\documentclass[amsmath,amssymb,superscriptaddress,showpacs]{revtex4}
\usepackage{graphicx}
\usepackage{dcolumn}
\usepackage{bm}
\usepackage[colorlinks=true, citecolor=blue, urlcolor = blue, linkcolor= red, bookmarks=true]{hyperref}

\begin{document}

\title{Generating black holes in $4D$ Einstein-Gauss-Bonnet gravity}
\author{Sushant G. Ghosh} \email{sghosh2@jmi.ac.in, sgghosh@gmail.com}
\affiliation{Centre for Theoretical Physics, Jamia Millia Islamia, New Delhi 110 025, India}
\affiliation{Astrophysics and Cosmology	Research Unit, School of Mathematics, Statistics and Computer Science, University of KwaZulu-Natal, Private Bag 54001, Durban 4000, South Africa}
\author{Rahul Kumar}\email{rahul.phy3@gmail.com}
\affiliation{Centre for Theoretical Physics, Jamia Millia Islamia, New Delhi 110 025, India}

\begin{abstract}
In recent times there is a surge of interest in constructing Einstein-Gauss-Bonnet (EGB) gravity, in the limit  $D \to 4 $,  of the $D$-dimensional EGB gravity.  Interestingly, the static spherically symmetric solutions in the various proposed $D \to 4 $  regularized EGB gravities coincide, and incidentally some other theories also admit the same solution.  We prove a theorem that characterizes a large family of nonstatic or radiating spherically symmetric solutions to the $4D$ EGB gravity, representing, in general, spherically symmetric Type II fluid.   An extension of the theorem, given without proof as being similar to the original theorem, generates static spherically symmetric black hole solutions of the theory.  It not only enables us to identify available known black hole solutions as particular cases but also to generate several new solutions of the $4D$ EGB gravity. 
\end{abstract}

%Use showkeys class option if keyword
                              
\maketitle

\section{Introduction}
Amongst, other solutions of general relativity (GR) or any modified theories of gravity, black holes remain one of the most exciting and active areas of study, since they throw challenging questions about the fundamental interactions between gravity and quantum mechanics. The concept of black holes began very shortly after Einstein's GR came to existence, Schwarzschild \cite{Schwarzschild:1916uq} found the solution to Einstein's equations in vacuum. Soon after, the electrovacuum static black hole solution  \cite{Reissner}, since then, numerous stationary black hole solutions sourced by some energy-matter distributions have been reported. Though the black hole's uniqueness theorems \cite{Israel1:1967wq} encapsulate that, in the Einstein-Maxwell theory, the unique black hole solutions are stationary and axially symmetric and three parameters defined them, however, in the presence of complicated matter fields distributions, uniqueness theorems or even black hole solutions are challenging to obtain.

Nevertheless, Salgado \cite{ms} proved a theorem characterizing a three-parameter family of static and spherically symmetric black hole solutions to Einstein equations by imposing certain conditions on the energy-momentum tensor (EMT).  This theorem allows the generation of a large family of exact static spherically symmetric black hole solutions, including their generalization to asymptotically de Sitter/Anti-de Sitter (dS/AdS) spacetimes. Salgado \cite{ms} work was promptly extended to higher-dimensional spacetime by Gallo \cite{gallogrg}. Though the static solutions should represent the eventually steady state of the dynamic evolution of black holes, this is not the most physical scenario, and one would like to consider dynamical black hole solutions, i.e., black holes with non-trivial time dependence.  However, due to the complexity of the Einstein field equations, such solutions are intractable and very few meaningful dynamical or nonstatic solutions are known. The Vaidya metric \cite{pc} is one of the nonstatic solution of Einstein's equations with spherical symmetry  whose metric in the Eddington-Finkelstein coordinates $\{v,r,\theta,\phi\}$ has a form \cite{pc},
\begin{equation}
ds^2 = - \left[1 - \frac{2 m(v)}{r}\right] d v^2 + 2 \epsilon d v
d r + r^2 (d \theta^2+ \sin^2 \theta d \phi^2), \hspace{0.5in}
\epsilon \pm 1 \label{vm}
\end{equation}
for a null fluid (radiation) source (a Type II fluid \cite{he}) described by EMT $T_{ab} = \psi l_a l_b$, $ l_a$ being a null vector field, and $m(v)$ is the mass function in advance time $v$. The Vaidya geometry permitting the incorporations of the effects of null fluid or null dust offers a more realistic background than static geometries. The Vaidya solution commonly used as an exterior solution for gravitational collapse models consisting of heat-conducting matter \cite{Herrera:1997ec} and useful to get insights in gravitational collapse situations \cite{Joshi:1987wg}, as a testing ground for the Cosmic Censorship Conjecture (CCC) \cite{Penrose:1964wq}, 
to model the dynamical evolution of a Hawking evaporating black holes \cite{rp1}, and in the stochastic gravity program \cite{hv}. Dawood and Ghosh \cite{Kothawala:2004fy} have extended Salgado's theorem \cite{ ms} which made it possible to generate nonstatic spherically symmetric Type II fluid that includes most of the known Vaidya solutions to Einstein field equations,  which was generalized by them for higher-dimensions \cite{sgad} and also to the higher-dimensional Einstein-Gauss-Bonnet (EGB) gravity \cite{gallo}. 

The Lovelock theories of gravity  \cite{Lovelock:1971yv}, the most natural possible generalizations of Einstein’s theory to higher dimensions,  are the only Lagrangian-based theories of gravity that give covariant, conserved, second-order field equations, and yields non-trivial dynamics in  $D\geq 5$.  The  EGB gravity \cite{Lanczos:1938sf} is a special case of Lovelock's theory of gravitation \cite{Lovelock:1971yv}, whose Lagrangian contains just the first three terms and is of particular interest. It appears naturally in the low energy effective action of heterotic string theory \cite{Gross}. Boulware and Deser \cite{bd} found exact black hole solutions in $D\; (\geq 5)$-dimensional  EGB gravitational theory. Later several interesting solutions were obtained to the EGB theory for various sources \cite{jtw,Wiltshire:1985us,egb}.

In four-dimensional ($4D$) spacetime, the Gauss-Bonnet term does not contribute to the gravitational dynamics since it becomes a total derivative.  Recently,  the EGB gravity theory reformulated in which the Gauss-Bonnet coupling has been re-scaled as $\alpha/(D-4)$ \cite{Glavan:2019inb} and $4D$  EGB theory obtained as the limit $D\to 4$ at the level of equations of motion.  The theory preserves the number of degrees of freedom and thereby free from the Ostrogradsky instability \cite{Glavan:2019inb}. Further, this natural extension of  Einstein's gravity bypasses all conditions of Lovelock's theorem \cite{Lovelock:1972vz}. The main idea here is to introduce a divergence that exactly cancels the vanishing contribution that the Gauss-Bonnet term makes to the field equations in $4D$.  

This stimulated research in $4D$ EGB gravity and various solutions of the theory have been found, namely the static spherically symmetric black holes \cite{Glavan:2019inb} and their charged extension \cite{Fernandes:2020rpa,Singh:2020nwo}, rotating black holes and their shadows \cite{Wei:2020ght,Kumar:2020owy}, Vaidya-like radiating black holes \cite{Ghosh:2020vpc},  noncommutative inspired black holes \cite{Ghosh:2020cob}, regular black holes \cite{Kumar:2020xvu,Kumar:2020uyz} and relativistic stars solution \cite{Doneva:2020ped}. Furthermore, the quasinormal modes, stability and shadows of spherically symmetric black holes \cite{Konoplya:2020bxa,Guo:2020zmf}, the motion of a classical spinning test particle \cite{Zhang:2020qew}, gravitational lensing \cite{Islam:2020xmy,Heydari-Fard:2020sib,Jin:2020emq,Kumar:2020sag},  derivation of regularized field equations \cite{Fernandes:2020nbq}  and thermodynamical phase transitions in AdS space \cite{Hegde:2020xlv,Singh:2020mty,Wei:2020poh,EslamPanah:2020hoj,HosseiniMansoori:2020yfj} have also been investigated. The extension to higher-order Lovelock gravity is presented in Refs.~\cite{Konoplya:2020qqh,Casalino:2020kbt}.

Tomozawa \cite{ Tomozawa:2011gp} originally initiated the discussion on the $4D$ regularization procedure of EGB gravity, and later  Cognola {\it et al.} \cite{Cognola:2013fva} simplified the approach by reformulating it, which mimics quantum corrections due to a Gauss-Bonnet invariant within a classical Lagrangian approach. The regularization procedure proposed in \cite{Glavan:2019inb} is at present a subject of dispute and number of question has been raised \cite{Ai:2020peo,Hennigar:2020lsl,Shu:2020cjw,Gurses:2020ofy,Mahapatra:2020rds} on the existence of $4D$ EGB gravity and also several alternate ways of the regularization have also been proposed \cite{Lu:2020iav,Kobayashi:2020wqy,Hennigar:2020lsl,Casalino:2020kbt,Ma:2020ufk,Arrechea:2020evj,Aoki:2020lig}.  Amongst these  L\"{u} and Pang  \cite{Lu:2020iav} regularized the $ 4D $  EGB gravity, via the Kaluza-Klein-like technique, by compactifying $D$ dimensional EGB gravity  on $D-4$ dimensional maximally symmetric  space, followed by redefining the coupling as $ \alpha /(D-4)$,  and then taking the limit $D \to 4$.  The technique leads to a  well defined and finite action of a special scalar-tensor theory that belongs to the family of Horndeski gravity,  in agreement with the results of Ref.~\cite{Kobayashi:2020wqy}.
This theory reproduced the spherically symmetric black hole obtained in Ref.~\cite{Glavan:2019inb} as a solution of trace equation for a particular scalar configuration unaffected by the curvature of the internal space \cite{Lu:2020iav}. Furthermore  while investigating  $2D$ Einstein gravity,  several distinct features were pointed out when applying the said approach  \cite{Ai:2020peo}, where one may also construct $2D$ black hole solutions \cite{Nojiri:2020tph}. After that, Hennigar  {\it et al.} \cite{Hennigar:2020lsl} proposed another well defined $D \to 4$ limit of EGB gravity generalizing the previous work of Mann and Ross in obtaining the $D \to 2$ limit of GR \cite{Mann} and this regularization is applicable not only in $4D$ but also in lower dimensions.  It was explicitly demonstrated by Hennigar  {\it et al.} \cite{Hennigar:2020lsl} that $ 4D$ spherically symmetric solutions  coincide with the solutions in \cite{Glavan:2019inb}, but is no longer true while finding more complicated geometry such as Taub-NUT solutions \cite{Hennigar:2020lsl}. 
Another alternate method of regularization for $4D$ EGB gravity  \cite{Fernandes:2020nbq}, also based on the Mann and Ross work \cite{Mann}, is proposed by adding counterterms, sufficient to cancel divergence of the action, yielding a set of field equations that can be written in closed form for $ 4D $ spacetime. This method is based on divergence-free action principle that belongs to the Horndeski gravity in general. It is completely independent of the compactification of higher dimensional spacetime to $4D$, but interestingly the trace equation could be made decouple from the scalar field and found to be identical to that of \cite{Glavan:2019inb,Cognola:2013fva}. It is also a scalar-tensor theory of the Horndeski type obtained by dimensional reduction methods. Thereby again, the maximally symmetric solutions of this theory are the same as in \cite{Glavan:2019inb}. Thus, we can conclude that the spherically symmetric $4D$  black hole solution obtained in  \cite{Glavan:2019inb} remains valid for these regularised theories \cite{Lu:2020iav,Hennigar:2020lsl,Casalino:2020kbt,Fernandes:2020nbq,Ma:2020ufk}.

The caveat in the regularization procedure of \cite{Glavan:2019inb} is that the limit $D\to 4$ at the field equations level leads to the unique solution only for spacetimes with enhanced symmetries, namely maximal or spherical symmetry. Indeed, this dimensional regularization procedure explicitly depends on the choice of higher-dimensional spacetime, and not every higher-dimensional solution allows for the $4D$ regularization, simply because there may be no $4D$ analogue of the corresponding higher-dimensional system. However, the alternate regularization procedures, leading to a divergence-free $4D$ action belonging to Horndeski gravity, make no \textit{a priori} assumptions about the symmetries of underlying geometry \cite{Fernandes:2020nbq,Hennigar:2020lsl} and the $4D$ regularized action is identical with that obtained from the Kaluza-Klein route \cite{Lu:2020iav,Kobayashi:2020wqy}. Besides, the black hole solution with more complicated geometry from the ($D\to 4$) limit of higher dimensional theory may not coincide with those obtained from these alternate regularized EGB theory \cite{Hennigar:2020fkv,Hennigar:2020lsl}.

The main aim of this work is to prove theorems in Refs.~\cite{sgad,gallo} so that a large family of exact spherically symmetric Type II fluid solutions are possible, including its generalization to asymptotically dS/AdS within the framework of the  $4D$ EGB gravity. As a result, we can find the analogous several GR solutions in the $4D$ EGB gravity. We firmly believe that our results are independent of the debate outcome and that they are also valid in the alternate $ 4D$ regularized EGB gravity theories.  Further, our spherically symmetric black hole solutions are valid in $4D$ non-relativistic Horava-Lifshitz theory of gravity \cite{Kehagias:2009is}, semi-classical Einstein's equations with conformal anomaly \cite{Cai:2009ua}, and gravity theory with quantum corrections \cite{Cognola:2013fva,Tomozawa:2011gp}. 

The paper is organized as follows: In section II, we review the $4D$ EGB gravity theory and recall the static spherically symmetric black hole solution. In section III, we prove the theorem for generating dynamical black holes in the $4D$ EGB gravity and discuss some particular cases. In section IV, we investigate the imposition of energy conditions on these metrics. Section V is devoted to the construction of static black holes. We summarize our findings and discuss possible future works in section VI.

\section{The $4D$ Einstein-Gauss-Bonnet gravity}\label{sec2} 
The action for EGB gravity theory reads 
\begin{equation}
S= \frac {1}{16\pi } \int d^Dx \sqrt {-g}  \Bigr[ R - 2\Lambda + {\alpha} (R_{a b c d } R^{a b c d } - 4R_{a b } R^{a b} + R^2 )\Bigl]+S_{\text{mat}},\label{action}
\end{equation}
where $R$, $R_{ab}$ and ${R}{^a}_{bcd}$ are, respectively, the Ricci scalar, Ricci tensor and Riemann curvature tensor, $g$ is the determinant of the metric tensor $g_{ab}$, $\Lambda$ is the cosmological constant, and $S_{\text{mat}}$ is the matter fields action. Varying action Eq.~(\ref{action}) with $g_{ab}$, i.e., $\delta S/{g^{ab}}=0$, yield the equations of motion as follow 
%\begin{widetext}
\begin{eqnarray}
8\pi T_{a b } &=& \mathcal{G}_{a b}=G^{(0)}_{a b}+G^{(1)}_{a
b}+G^{(2)}_{a b},\label{FieldEq}
\end{eqnarray}
where $T_{a b}$ is the EMT associated with the matter-field distribution resulting from the variation $\delta
S_{\text{mat}}/\delta g^{a b},$ and
\begin{eqnarray}
G^{(0)}_{a
b}&=&\Lambda g_{a b}  \\
 G^{(1)}_{a
b}&=& R_{a b }-\frac{1}{2}Rg_{a b}\\
G^{(2)}_{a b}&=&-{\alpha} \Bigr[\frac{1}{2} g_{a b} (R_{c j e k}R^{c j e k }-4R_{c j}R^{c j} +R^2)\nonumber  \\
&-& 2RR_{a b}+4R_{a c}R^{c}_{b}+4R_{a c b j} R^{c j}-2R_{a c j e}R_{b}^{ \ c j e} \Bigl],
\end{eqnarray}
where $G^{(1)}_{a b}$ and $G^{(2)}_{a b}$, respectively, are the Einstein's tensor and the Lanczos's tensor \cite{Lanczos:1938sf}. The $4D$ theory is defined by rescaling the Gauss-Bonnet coupling constant $\alpha/(D-4)$ and taking  limit as $D\to 4$, at the level of equations of motion rather than at the level of the action \cite{Glavan:2019inb}.

Taking the $D$ dimensional static and spherically symmetric metric  
\begin{equation}
ds^2=-F(r)dt^2+\frac{1}{F(r)}dr^2
+r^2d\Omega^2_{D-2} \; ,
\end{equation}
with 
\begin{equation}
d\Omega^2_{D-2} = d\theta^2_1 + \sum^{D-2}_{i=2}\prod^{i-1}_{j=1}
\sin^{2}\theta_j\;d\theta^2_i \; ,
\end{equation}
as ansatz, and solving the field equation (\ref{FieldEq}) in the limit $D\to 4$, we get the static spherically symmetric black hole solution in the   $4D$ EGB gravity with  \cite{Glavan:2019inb,Fernandes:2020rpa}
\begin{equation}
F(r)=1+\frac{r^2}{2\alpha}\left(1\pm\sqrt{1+\frac{8M\alpha}{r^3}+\frac{4\Lambda\alpha}{3}}\right).\label{fr}
\end{equation}
Here, $M$ is the black hole mass and the two branches of solutions corresponds for the $``\pm"$ sign. At large distances, Eq.~(\ref{fr}) reduces to 
\begin{eqnarray}
F_-(r)=1-\frac{2M}{r\sqrt{1+\frac{4\Lambda\alpha}{3}}} +\frac{r^2}{2\alpha}\Bigr(1-\sqrt{1+\frac{4\Lambda\alpha}{3}}\Bigl)+\mathcal{O}\Big(\frac{1}{r^3}\Big),\nonumber\\
F_+(r)=1+\frac{2M}{r\sqrt{1+\frac{4\Lambda\alpha}{3}}} +\frac{r^2}{2\alpha}\Bigr(1+\sqrt{1+\frac{4\Lambda\alpha}{3}}\Bigl)+\mathcal{O}\Big(\frac{1}{r^3}\Big),
\end{eqnarray} 
which shows that only the $``-"$ branch solution, $F_{-}(r)$, that asymptotically goes over to the Schwarzschild black hole with the correct mass sign, i.e., has the correct limit for $\alpha\rightarrow 0$. We consider that the Gauss-Bonnet coupling parameters $\alpha$ is positive and $\Lambda$ is a constant quantity. Though the semi-classical gravity with a conformal anomaly \cite{Cai:2009ua} and the theory of gravity with quantum corrections \cite{Cognola:2013fva} also admit similar black hole solutions as found in Eq.~(\ref{fr}), the EGB gravity can be considered as a classical modified gravity theory on equal footing with GR. Regularized Lovelock gravity with an arbitrary curvature order, when truncated at quadratic order, also admits the similar black hole solution \cite{Casalino:2020kbt}. 

Next, we consider a theorem, so that a large family of exact spherically symmetric dynamical black hole solutions, for the $4D$ EGB gravity, are possible. The generated solutions represent a generalization of  Vaidya-like solutions to this theory. 

\section{Radiating black holes solutions}\label{sec3}
\textbf{Theorem I:} \textit{Let $(\mathcal{M},g_{ab})$ be a
$D$-dimensional space-time such that: i) it satisfies in the limit $D \to 4 $ the Einstein-Gauss-Bonnet gravity equations obtained by re-scaled coupling constant $\alpha/(D-4)$, ii) it is spherically symmetric, iii) in the Eddington-Bondi coordinates, where the metric reads $\,ds^2=-A^2(v,r)F(v,r)dv^2+2\epsilon A(v,r)dvdr+r^2d\Omega^2_{D-2}$, the EMT $T^a_b$ satisfies the conditions	$T^v_r=0$, and $T^{\theta_1}_{\theta_1}=\gamma T^r_r$, $(\gamma =\text{const}\in \mathbb{R})$, iv) if $\alpha\rightarrow 0$, the solution converges to the $4D$ GR limit. Then the metric of the spacetime is given by
\begin{eqnarray} ds^2=-F(v,r)dv^2+2\epsilon
dvdr+r^2d\Omega^2_2,\,~~~~~~(\epsilon=\pm1),\label{metrica}\end{eqnarray}
where
\begin{widetext}
	\begin{equation}
F_{\pm}(v,r)=\left\{
	\begin{array}{rll}
	1+ \frac{r^2}{2\alpha}\left \{1\pm \sqrt{1+\frac{4\Lambda\alpha}{3} +\frac{8M(v)\alpha}{r^3}-\frac{32\pi\alpha 
			C(v)}{(1+2\gamma )r^{2(1-\gamma )}}}\right \}\;&\text{if}&\;\; \gamma \neq-\frac{1}{2},\\
	1+ \frac{r^2}{2\alpha}\left \{1\pm \sqrt{1+\frac{4\Lambda\alpha}{3} +\frac{8M(v)\alpha}{r^3}-\frac{32\pi\alpha 
			C(v)\ln r}{r^3}}\right \}\;&\text{if}&\;\; \gamma =-\frac{1}{2},\\
	\end{array}\\\right.\label{lametrica}
	\end{equation}
\end{widetext}
with the diagonal components of EMT $T^a_b$ given by 
\begin{equation}
T^a_{b\text{(Diag)}}=\frac{C(v)}{r^{2(1-\gamma)}}\text{diag}[1,1,\gamma ,\gamma],\label{Trr}
\end{equation}
and only non-vanishing off-diagonal element as
\begin{equation} T^r_v=\left\{
\begin{array}{lll}
\frac{1}{4\pi
r^{2}}\frac{dM(v)}{dv}-\frac{r^{2\gamma -1}}{2\gamma +1}\frac{dC(v)}{dv}\;\;
&\text{if}&\;\; \gamma \neq -\frac{1}{2},\\
\frac{1}{4\pi
r^2}\frac{dM(v)}{dv}-\frac{\ln(r)}{r^2}\frac{dC(v)}{dv} \;\;
&\text{if}&\;\; \gamma = -\frac{1}{2},\\
\end{array}\\\right.
\end{equation}
where $M(v)$ and $C(v)$ are two arbitrary functions depending on the distribution of the underlying matter.}

\textbf{Proof:} By the hypothesis iii) of the Theorem I, we start with the metric for higher dimensional spherically symmetric spacetime in Eddington coordinates
\begin{equation}
ds^2=-A^2(r,v)F(v,r)dv^2+2\epsilon A(r,v)dvdr+r^2d\Omega^2_{D-2},\label{anstaz}
\end{equation}
where $\epsilon=-1, +1$, respectively, correspond to the outgoing and ingoing null fluid. Due to the hypothesis i), metric (\ref{anstaz}) must satisfy the EGB field equations (\ref{FieldEq}). Considering the special case $T^v_r=0$ (hypothesis (iii)), Eq.~(\ref{FieldEq}) yields
\begin{equation}\label{Gvr}
\mathcal{G}^{v}_r= (D-2)\Big[r^2+2(D-3)\alpha(1-F)\Big]\frac{1}{r^3\epsilon A^2}\Bigr(\frac{\partial A}{\partial r}\Bigl),
\end{equation}
which implies that in the limit $D \to 4$, Eq.~(\ref{Gvr}) solves to $A(v,r)=g(v)$. However, re-defining the null coordinate as $\overline{v} = \int g(v) dv$, we can always set, without the loss of generality, $A(v,r) = 1$.

Now, from the $(r,\;r)$ and $(v,\;v)$ components of the field equations (\ref{FieldEq}), we obtain that
$\mathcal{G}^v_v=\mathcal{G}^r_r$, which further ensure that $$T^v_v=T^r_r.$$
Thus the EMT can be written as :
\[
T^a_b = \left(%
\begin{array}{cccccc}
T^v_v \; & T^v_r\;  & 0 \;  & 0 \;  & . \;  & . \; \\

T^r_v & T^r_r & 0 & 0  & . \;  & . \; \\
0 & 0 &T^{\theta_1}_{\theta_1}& 0  & . \;  & . \; \\
. & . & . & .    & . \;& . \; \\
. & . & . & .    & . \;&  T^{\theta_{D-2}}_{\theta_{D-2}}\; \\
\end{array}%
\right),
\]
which in general belongs to a Type II fluid with
$T^{\theta_1}_{\theta_1}=T^{\theta_2}_{\theta_2} =\, .\, .\, .\, =
T^{\theta_{D-2}}_{\theta_{D-2}}$. It may be worthwhile to recalled that the EMT of a Type II fluid has a double null eigenvector, whereas that for a
Type I fluid has only one time-like eigenvector \cite{he}.
If we impose the conservation laws, $\nabla_a T^a_b=0,$ and using
again the hypothesis $iii)$, $T^{\theta_i}_{\theta_i}=\gamma T^r_r$, we
have that
\begin{eqnarray}
& &\frac{\partial }{\partial v}{ T^v_v} +\frac {\partial
}{\partial r} T^r_v +\frac{1}{2\epsilon}\left (T^r_r- T^v_v\right
){\frac
{\partial }{\partial r}F}+ \frac {( D-2)}{r}T^r_v =0 ,\label{Bt11}\\
& &\frac {\partial }{\partial r}T^r_r +\frac {( D-2)(1-\gamma)}
{r}T^r_r =0. \label{Bt22}
\end{eqnarray}
Solving Eq.~(\ref{Bt22}) for $T^r_r$, we obtain:
\begin{equation}
T^r_r=\frac{C(v)}{r^{(D-2)(1-\gamma)}},\label{Trr}
\end{equation}
where $C(v)$ is an arbitrary function. Using these results, the diagonal elements of $T^a_b$ can be written as follow
$$ T^a_{b\text{(Diag)}}=\frac{C(v)}{r^{(D-2)(1-\gamma)}}\text{diag}[1,1,\gamma ,\cdots,\gamma].$$ Now, in the limit $D\to 4$, the EGB equations $\mathcal{G}^r_r=8\pi T^r_r$, reduces to

\begin{table*}
	\caption{\label{tab:table3} Some spacetimes generated from the {\bf Theorem I }for particular
		values of $\gamma $, $C(v)$ and $M(v).$ Those with references are  known solutions, and others are new.}
	\begin{ruledtabular}
		\begin{tabular}{llll}
			
			$T^a_b$&Space-Time&$M(v)$ and $C(v)$ &$\gamma$ -index\\ \hline\\
			$T^a_{b}=0\;,\;T^r_v=\frac{1}{4\pi r^2}\frac{dM(v)}{dv}$&Vaidya-$4D$ EGB \cite{Ghosh:2020vpc}&$M(v),\;C(v)=0$
			&$\gamma =0$\\\\\hline\\
			$T^a_{b}=-\frac{Q^2(v)}{8\pi r^4}\text{diag}[1,1,-1,-1]$&Bonnor-Vaidya-$4D$ EGB \cite{Ghosh:2020vpc}
			&$M(v),\;C(v)=-\frac{Q^2(v)}{8\pi}$&$\gamma =-1$\\\\ $T^r_v=\frac{1}{4\pi r^3} \left[r\frac{dM(v)}{dv}-Q(v)\frac{dQ(v)}{dv}\right]$&&&
			\\\\\hline\\
			$T^a_{b}=-\frac{g^2(v)}{4\pi r^{2(m+1)}}\text{diag}[1,1,-m,-m]$&Husain-$4D$ EGB&$M(v),\;C(v)=
			-\frac{g^2(v)}{4\pi}$&$\gamma =-m$\\\\
			$T^r_v=\frac{1}{4\pi r^3} \left[r\frac{dM(v)}{dv}+\frac{2g(v)r^{2(1-m)}}{ (1-2m)}\frac{dg(v)}{dv}\right]$
				\\\\\hline\\
			$T^a_{b}=-\frac{(1+2\gamma)q(v)}{8\pi r^{2(1-\gamma)}}\text{diag}[1,1,\gamma,\gamma]$&Radiating quintessence-$4D$ EGB&$M(v),C(v)=
			-\frac{(1+2\gamma)q(v)}{8\pi}$&$0<\gamma<1$\\\\
			$T^r_v=\frac{1}{4\pi r^3} \left[r\frac{dM(v)}{dv}+\frac{r^{2(\gamma+1)}}{ 8\pi}\frac{dq(v)}{dv}\right]$&&&
			\\\\\hline\\
			$ T^v_v=T^r_r=-\frac{a}{4\pi r^2}$& Clouds of strings / global monopole-$4D$ EGB \cite{Singh:2020nwo}
			&$M(v)=M, \;\;C(v)=-\frac{a}{4\pi}$&$\gamma =0$\\\\\hline\\
			$T^a_b=0$& Glavan-Lin \cite{Glavan:2019inb}
			&$M(v)=M$, $C(v)=0$&$\gamma =0$			\\\\\hline\\
			$T^a_{b}=-\frac{Q^2}{8\pi r^4}\text{diag}[1,1,-1,-1]$& Fernandes \cite{Fernandes:2020rpa}
			&$M(v)=M,\,C(v)=-\frac{Q^2}{8\pi}$&$\gamma =-1$\\\\
			$T^r_v=0$&&&	\\\\\hline\\
			$T^a_b=0$& dS/AdS
			&$M(v)=0$, $C(v)=0$&$\gamma =0$		\\\\\hline\\
	 	$T^a_{b}=-\frac{(1+2\gamma)q}{8\pi r^{2(1-\gamma)}}\text{diag}[1,1,\gamma,\gamma]$&Quintessence-$4D$ EGB&$M(v)=M,\;C(v)=
		-\frac{(1+2\gamma)q}{8\pi}$&$0<\gamma<1$
		\\\\
		\end{tabular}
	\end{ruledtabular}
\end{table*}

\begin{equation}
\Lambda+\Big[r^2+2\alpha (1-F)\Big] \frac{1}{r^3}\frac{\partial F}{\partial r} -\Big[r^2-\alpha (1-F)\Big] \frac{1-F}{r^4}=\frac{8\pi C(v)}{r^{2(1-\gamma )}}.\label{rreq}
\end{equation}
After solving this differential equation and making some
algebraic simplifications, we get
\begin{widetext}
\begin{equation}
F_{\pm}(r,v)=\left\{
\begin{array}{rll}
  1+ \frac{r^2}{2\alpha}\left \{1\pm \sqrt{1+\frac{4\Lambda\alpha}{3} +\frac{8M(v)\alpha}{r^3}-\frac{32\pi\alpha 
C(v)}{(1+2\gamma)r^{2(1-\gamma)}}}\right \}\;&\text{if}&\;\; \gamma \neq-\frac{1}{2},\\
 1+ \frac{r^2}{2\alpha}\left \{1\pm \sqrt{1+\frac{4\Lambda\alpha}{3} +\frac{8M(v)\alpha}{r^3}-\frac{32\pi\alpha 
 		C(v)\ln r}{r^3}}\right \}\;&\text{if}&\;\; \gamma =-\frac{1}{2}.\\
\end{array}\\\right.\label{soln}
\end{equation}
\end{widetext}
where $M(v)$ is another arbitrary function, which can be identified as the mass of the underlying matter. 
Finally to calculate the only non-zero off-diagonal component $T^r_v$. From the EGB equation, we obtain
$$\mathcal{G}^r_v=8\pi T^r_v,$$
we find that the only non-vanishing off-diagonal element of
$T^a_b,$ reads
\begin{equation}
T^r_v=\left\{
\begin{array}{lll}
\frac{1}{4\pi
r^2}\frac{dM(v)}{dv}-\frac{r^{2\gamma -1}}{1+2\gamma}\frac{dC(v)}{dv}\;\;
&\text{if}&\;\; \gamma \neq -\frac{1}{2},\\
\frac{1}{4\pi
r^2}\frac{dM(v)}{dv}-\frac{\ln(r)}{r^2}\frac{dC(v)}{dv} \;\;
&\text{if}&\;\; \gamma = -\frac{1}{2}.\\
\end{array}\\\right.
\end{equation}
It is seen that Eq.~(\ref{Bt11}) is identically satisfied and  the Theorem is proved. \\

Some critical comments are in order. The result of the {\bf Theorem I} represents a general class of non-static, spherically symmetric solutions to the   $4D$ EGB theory describing radiating black-holes with the EMT satisfying the conditions per the hypothesis (iii).
In general, the family of the solutions outlined by the {\bf Theorem I } generates solutions of the   $4D$ EGB theory, for instance,
Bonnor-Vaidya-like \cite{Ghosh:2020vpc}, dS/AdS \cite{ww}, global monopole-like \cite{Barriola:1989hx}, Husain \cite{Husain:1995bf}  and  Dadhich-Ghosh Vaidya solution on brane-like \cite{dg}.  Clearly, by proper choice of the functions $M(v)$ and $C(v)$, and $\gamma-$index, one can generate solutions of the theory, and some of them are presented in the Table \ref{tab:table3}. They include most of the known Vaidya-based spherically symmetric solutions of the   $4D$ EGB theory. These solutions could be beneficial to study the collapse of different matter fields or the formation of naked singularities.
Furthermore, these solutions can be used to get some insights into the semi-classical approaches for black holes evaporation. \\ 

To further illustrate the theorem, we generate the two known solutions of the   $4D$ EGB theory.
\paragraph{Glavan and Lin solution \cite{Glavan:2019inb}:}
As an immediate consequence of the theorem, we generate the static spherically symmetric black hole solution of the   $4D$ EGB theory \cite{Glavan:2019inb}.  For this we have to choose, $M(v)\equiv M=constant$, $C(v)=0$, and $\Lambda=0$, solution~(\ref{soln}), becomes static and is given by 
\begin{equation}\label{gl}
F_{\pm}(r)=1+\frac{r^2}{2\alpha}\left(1\pm\sqrt{1+\frac{8M\alpha}{r^3}}\right). 
\end{equation}
The metric (\ref{anstaz}) with $F(r)$ in (\ref{gl}) represents the static spherically symmetric black hole solution obtained by Glavan and Lin \cite{Glavan:2019inb}, but in the Einstein-Finkelstein coordinates.
\paragraph{Fernandes solution \cite{Fernandes:2020rpa}:} The charged counterpart of the spherically symmetric black hole of the   $4D$ EGB theory \cite{Glavan:2019inb} were also found by Fernandes \cite{Fernandes:2020rpa}.  To generate this, we should choose   $M(v)\equiv M=constant$,  $C(v)=-Q^2/8\pi$, $\Lambda= -3/l^2$ and $\gamma =-1$. 

\paragraph{GR limit:}
At this point, it is important to note that we have obtained two branches of solution (\ref{soln}), namely, $F_+$ and $F_-$, which correspond to $\pm$ signs
in front of the square root term. However, the positive branch, $F_+$, does not converge to GR, henceforth, we will be considering only negative branch. In this case, the limit $\alpha\rightarrow 0$, reduces $F(v,r)$ to
\begin{widetext}
	\begin{equation}
	F_-(v,r)=\left\{
	\begin{array}{lll}
	1-\frac{2M(v)}{r}-\frac{\Lambda r^2}{3} +\frac{8\pi
		C(v)r^{2\gamma}}{(1+2\gamma)}\;&\text{if}&\;\; \gamma \neq-\frac{1}{2},\\\\
	1-\frac{2M(v)}{r}-\frac{\Lambda r^2}{3} +\frac{8\pi
		C(v)\ln r}{r};&\text{if}&\;\; \gamma =-\frac{1}{2}.\\\\
	\end{array}\\\right.
	\end{equation}
\end{widetext}
These are solutions for the $4D$ GR version of the Theorem, i.e., solution of $G^{(0)}_{a b}+G^{(1)}_{a b}=8\pi T_{a b}$, instead of the EGB gravity, which are the same as those found in Ref.~\cite{sgad}. 

\section{Energy conditions}\label{sec4}
The family of solutions discussed here, in general, belongs to Type II fluid defined in Ref.~\cite{he}. To discuss the energy conditions, let us introduce two independent future null vectors, $l_a$ and $n_a$, where $l^a$ is tangent to the null surface constructed by $v$, and $n^a$ is an another independent null vector such  that 
\begin{eqnarray}
\l_{a} &=& - \delta_a^v, \: n_{a} = - \frac{1}{2} F(v,r)
\delta_{a}^v + \delta_a^r, \label{nvagb} \\
l_{a}l^{a} &=& n_{a} n^{a} = 0, \; ~l_a n^a = -1,\; ~l^a n_a =1,
\label{nvdgb}
\end{eqnarray}
and the  EMT, with the help of
these null vectors, reads \cite{ww,he}
\begin{equation}
T_{ab}=\epsilon\; \mu (v,r)
l_al_b-P_r(v,r)(l_an_b+l_bn_a)+P_{\theta}(v,r)(g_{ab}+l_an_b+l_bn_a),\label{EMT}
\end{equation}
where
\begin{eqnarray}
\mu&=&T^r_v,\\
P_r&=&T^r_r=C(v)r^{2(\gamma -1)},\\
P_{\theta}&=&\gamma P_r,
\end{eqnarray}
where $\mu$ corresponds for the radiating energy along the null direction $l^a$; $P_r$ and $P_{\theta}$, respectively, are the radial and transverse pressures components generated by the charges of the fluids. All these physical quantities are measured in the reference frame of an observer moving along a time-like direction $u^a$ given by
$$u^a=\frac{1}{\sqrt{2}}(l^a+n^a).$$ 
The energy density $\rho$ measured by this observer is defined by the projection of $T_{ab}$ along the $u^a$, as follow
$$\rho=-T_{ab}u^au^b=-P_r.$$ 

\textit{i). The weak energy condition (WEC):} For any timelike vector $w^a$, the EMT $T_{ab}w^aw^b\geq 0$ \cite{ww,he}. Equation (\ref{EMT}) can be recast as 
\begin{equation}
\mu\geq 0,\;\;\; \rho\geq0 \;\;\; \text{and} \;\;\;\;P_{\theta}\geq 0.\label{wec}
\end{equation}
The WEC and strong energy condition (SEC) are identical for the Type II fluid \cite{sgad,ww,he}. 

\textit{ii). The dominant energy condition (DEC):} For any timelike vector $w^a$, $T_{ab}w^aw^b\geq 0$ and also
$T_{ab}w^b$ is a non-spacelike vector, i.e.,
\begin{equation}
\mu\geq 0\;\;\;\; \text{and} \;\;\;\;\rho\geq P_{\theta}\geq 0.\label{dec}
\end{equation}
For the radiating fluid, the WEC condition (\ref{wec}) is satisfied if $C(v)\leq 0$ and $\gamma\leq 0$. However, $\mu>0$, leads to
\begin{equation}
\frac{1}{4\pi
	r^2}\frac{dM(v)}{dv}-\frac{r^{2\gamma -1}}{2\gamma +1}\frac{dC(v)}{dv}>
0\;\; \text{if}\;\; \gamma \neq
-\frac{1}{2},\label{conkdis}\end{equation} and
\begin{equation}
\frac{1}{4\pi
	r^2}\frac{dM(v)}{dv}-\frac{\ln(r)}{r^2}\frac{dC(v)}{dv}> 0
\;\;\text{if}\;\; \gamma = -\frac{1}{2}, \label{conkigual}
\end{equation}
which is satisfied if $dM(v)/dv > 0$, and either ${dC(v)}/{dv}>0$ with $\gamma < - {1}/{2},$, or
${dC(v)}/{dv}<0$ with $\gamma > - {1}/{2}.$
On the other hand, the Eq.~(\ref{conkigual}) is satisfied if
\[\frac{dM(v)}{dv}>4\pi \ln(r)\frac{dC}{dv}.\]  Finally, the DEC conditions (\ref{dec}), $\rho\geq P_{\theta}\geq 0$, are satisfied only if $C(v)\leq 0$ and $-1\leq \gamma  \leq 0$.

\section{Static black holes solutions}\label{sec5}
The {\bf Theorem I} generates a general class of non-static, spherically symmetric solutions to the $4D$ EGB gravity representing radiating black holes with the EMT, which satisfies the conditions as per hypothesis (iii). One can also generate the static solutions in the Eddington-Finkelstein coordinates by setting $M(v) = M, \; C(v) = C$,
with $M$ and $C$ as constants, in which case matter is Type I. 
Then metric (\ref{anstaz}) can be transformed 
in the usual spherically symmetric form by the transformation 
\begin{equation}
ds^2 = -F(r)\; dt^2 + \frac{dr^2}{F(r)} + r^2 (d \Omega_{D-2})^2,
\end{equation}
by the coordinate transformation
\begin{equation}
dv = A(r)^{-1} \left( dt + \epsilon \frac{dr}{F(r)} \right).
\end{equation}
In case of spherical symmetry, even when $F(r)$ is replaced by $F(t,r)$, one can cast the metric in the form (\ref{anstaz})
\cite{Nielsen:2005af}.
Thus, one would
like to have the above theorem to generate static
spherically symmetric solutions which we state without proof. 

\textbf{Theorem II:} \textit{Let $(\mathcal{M},g_{ab})$ be a
	$D$-dimensional spacetime such that: i) it satisfies, $D \to 4 $, the Einstein-Gauss-Bonnet gravity equations obtained by re-scaled coupling constant $\alpha/(D-4)$, ii) it is spherically symmetric, iii) in the \textbf{spherical polar} coordinates, where the metric reads
	$\,ds^2=-F(r)dt^2+N(r) dr^2+r^2d\Omega^2_{D-2}$,
	the EMT $T^a_b$ satisfies the conditions
	$T^v_r=0$, and $T^{\theta_1}_{\theta_1}=\gamma T^r_r$,
	$(\gamma =\text{const}\in \mathbb{R})$, iv) if $\alpha\rightarrow 0$, the
	solution converges to the $4D$ GR limit. Then the metric of the spacetime is given by
	\begin{eqnarray} ds^2=-F(r)dt^2+\frac{1}{F(r)}dr^2+r^2d\Omega^2_2,\,~~~~~~\label{metrica}\end{eqnarray}
	where
\begin{widetext}
\begin{equation}
	F_{\pm}(r)=\left\{
	\begin{array}{rll}
		1+ \frac{r^2}{2\alpha}\left \{1\pm \sqrt{1+\frac{4\Lambda\alpha}{3} +\frac{8M\alpha}{r^3}-\frac{32\pi\alpha 
				C}{(1+2\gamma)r^{2(1-\gamma)}}}\right \}\;&\text{if}&\;\; \gamma \neq-\frac{1}{2},\\
		1+ \frac{r^2}{2\alpha}\left \{1\pm \sqrt{1+\frac{4\Lambda\alpha}{3} +\frac{8M\alpha}{r^3}-\frac{32\pi\alpha 
				C\ln r}{r^3}}\right \}\;\;\;\;\;&\text{if}&\;\; \gamma =-\frac{1}{2},\\
	\end{array}\\\right.\label{staticsoln}
\end{equation}
\end{widetext}
where $N(r) =1/F(r)$ and 
the components of EMT $T^a_b$ given by 
\begin{equation}
T^a_{b}=\frac{C}{r^{2(1-\gamma)}}\text{diag}[1,1,\gamma ,\gamma],\label{Trr}
\end{equation}
where $M$ and $C$ are two arbitrary constant depending on the distribution of the underlying matter.} \\

The {\bf Theorem II }generates a general class of static,
spherically symmetric black hole solutions to the   $4D$ EGB theory with the EMT, which satisfies the conditions as
per the hypothesis (iii).  The family of solutions outlined here contains
the   $4D$ EGB version, for instance,  of Glavan-Lin ~\cite{Glavan:2019inb} static spherically symmetric black hole when $C=0,\Lambda=0$ and  charged counterpart of spherically symmetric AdS  black hole dues to Fernandes \cite{Fernandes:2020rpa} by choosing $C(v)=-Q^2/8\pi$, $\Lambda= -3/l^2$ and $\gamma =-1.$

Obviously, by proper choice of the constant $M$ and $C$, and $\gamma-$index, one can generate as
many solutions as required. The above {\bf Theorem II } can generate several spherically symmetric solutions to the   $4D$ EGB theory with the EMT satisfying conditions mentioned in the theorem.  

\section{Conclusions}\label{sec6}
Lately, significant attention was devoted to several regularisations of EGB gravity to $4D$  after a proposal of defining $4D$ EGB  theory by rescaling Gauss-Bonnet coupling constant as  $\alpha/(D-4)$  and taking the limit $D \to 4$ of the $D$-dimensional solutions of EGB gravity.  The spherically symmetric black hole solution of the formulated $4D$ EGB gravity \cite{Glavan:2019inb, Cognola:2013fva} in contrast to the  Schwarzschild black hole solution of GR is free from the singularity pathology as the gravity becomes repulsive at short distances.   It is argued, without proof, that a physical observer could never reach this curvature singularity given the repulsive effect of gravity at short distances \cite{Glavan:2019inb}. However, later a geodesic analysis contradicts this observation about the singularity being unreachable by any observer in finite proper time \cite{Arrechea:2020evj}.   They explicitly showed that an infalling particle starts at rest will reach the singularity with zero velocity as attractive and repulsive effects compensate each other along the trajectory of the particle  \cite{Arrechea:2020evj}. Also, the   spherically symmetric black hole solution of other  $4D$ regularized theories \cite{Hennigar:2020lsl,Casalino:2020kbt,Lu:2020iav,Ma:2020ufk,Tomozawa:2011gp} coincides with that obtained for  the $4D$ EGB theory in Ref.~\cite{Glavan:2019inb,Cognola:2013fva}.

Whilst for finding the exact solutions of Einstein equations in the $4D$ spacetime several powerful mathematical tools developed,  it would be interesting how to develop some of these methods to get exact solutions of the more complicated higher curvature EGB gravity.  With this motivation, we have proved a theorem, which, with certain restrictions on the EMT characterizes a large family of radiating black hole solutions to this   $4D $ EGB gravity, representing, in general, spherically symmetric Type II fluid.  The solutions depend on one parameter $\gamma$, and two arbitrary functions $M(v)$ and $C(v)$ (modulo energy conditions). It is easy to generate various solutions by suitable choice of these functions and the parameter $\gamma$. In particular,  we have demonstrated that the known solutions of the theory are generated as the particular case using {\bf Theorem I},  and we have also listed some other solutions in the table \ref{tab:table3}, which means that there exist realistic matter that follows the restrictions of the theorem. Whilst,  we have generated a set of solutions of the   $4D $ EGB gravity, it is always desirable to see if there exist physically reasonable new solutions to extend this list.  

The family of solutions generated by the  {\bf Theorem I}, in general, belongs to Type II fluid. However, if $M(v) = C(v) = $ constant,  the matter field degenerates to a Type I matter with no off-diagonal component of the EMT, and one can generate static black hole solutions in the Eddington-Finkelstein coordinates, of the $4D$ EGB gravity, with appropriate choices of $M,\;C$ and $\gamma$. A trivial extension of the {\bf Theorem I} also stated as {\bf Theorem II}, without proof, is similar to that of {\bf Theorem I}, which allows one to generate a three-parameter family of static, spherically symmetric solutions of the $4D$ EGB gravity in the Schwarzschild coordinates. 

The $4D$ EGB  solution Eq.~(\ref{gl}), obtained in Ref.~\cite{Glavan:2019inb}, apart from the other regularized $4D$ theories, actually was also found earlier in the gravity with a conformal anomaly \cite{Cai:2009ua}, the $4D$ non-relativistic Horava-Lifshitz theory of gravity \cite{Kehagias:2009is}, and also recently in the Lovelock gravity \cite{Casalino:2020kbt,Konoplya:2020qqh}. Hence, the theorems presented here, with appropriate modifications, may also be relevant in these theories, and one can generate a family of both static and dynamical spherically symmetric solutions of these theories. 

Many interesting avenues are amenable for future work from the solutions generated; it will be intriguing to analyze the causal structure and thermodynamics. Also, it should be interesting to apply these metrics to study the effects of the higher-order curvature in a semi-classical analysis of the black hole evaporation in $4D$. One should also see the possibility of generalization of these results to more general Lovelock gravity theories. Further, the presented solutions may provide an excellent setting to get insights into more general gravitational collapse situations and in better understanding of CCC \cite{rp}.  

\appendix
\section{Quintessence $4D$ EGB black hole solution by alternate regularization techniques \cite{Hennigar:2020lsl,Lu:2020iav,Kobayashi:2020wqy}}
We have generated $4D$ EGB black hole solution surrounded by quintessence using our theorem (see Table \ref{tab:table3}). To further strengthen our claim, we derive this solution by alternate regularization techniques \cite{Hennigar:2020lsl,Lu:2020iav,Kobayashi:2020wqy}, and check if we get the same solution as obtained using the {\bf Theorem I}.   
An alternate $4D$ regularization procedure for EGB gravity is proposed via the Kaluza-Klein-like route of compactifying the $D$-dimensional EGB gravity on a $(D-4)$-dimensional maximally symmetric space \cite{Lu:2020iav,Kobayashi:2020wqy}. This leads to a well defined and divergence free action in $4D$ describing a scalar-tensor theory of gravity that belongs to a class of Horndeski gravity. Following \cite{Lu:2020iav}, we start with the $D$-dimensional EGB gravitational action (\ref{action}) and consider a Kaluza-Klein ansatz
\begin{equation}
	ds_D^2=ds_p^2+\exp[2\psi]d\Sigma^2_{D-p},
\end{equation}
where $d\Sigma^2_{D-p}$ is the line element on the internal maximally symmetric space of curvature proportional to $\lambda$, $ds_p^2$ is the $p$-dimensional line element, and scalar field $\psi$ is a function of external $p$ dimensional space coordinates. Redefining the Gauss-Bonnet coupling as $\alpha\to \alpha/(D-p)$ and taking the limit $D\to p $ in (\ref{action}), we obtained the $p$-dimensional reduced EGB gravitational action, which for $p=4$ reads as
\begin{align}
	S_4=&\frac {1}{16\pi }\int d^4x\sqrt{-g}\Big[R
	-2\Lambda+\alpha\Big(\psi\,\mathcal{L}_{GB}+4G^{\mu\nu}\partial_\mu\psi\partial_\nu\psi-2\lambda R e^{-2\psi} -4(\partial\psi)^2\Box \psi+2\left((\partial\psi)^2\right)^2\nonumber\\
	&-12\lambda(\partial\psi)^2e^{-2\psi}-6\lambda^2e^{-4\psi} -2(1+2\gamma) q r^{2 \gamma}   \Big)\Big],\label{action2}
\end{align}
and corresponds to the $4D$ regularized EGB gravity action with rescaled Gauss-Bonnet coupling constant. The action (\ref{action2}) is similar to that obtained in \cite{Hennigar:2020lsl} with trivial field redefinition and $\lambda=0$. Here, $\mathcal{L}_{GB}$ is Gauss-Bonnet Lagrangian density and $q$ is the quintessence parameter and $0<\gamma<1$. One can obtain the covariant field equations by varying the action (\ref{action2}) for metric tensor $g_{\mu\nu}$ and scalar field $\psi(r)$ \cite{Hennigar:2020lsl,Lu:2020iav}. To study the static spherically symmetric black hole solution, we consider the metric \textit{ansatz} and scalar field as follows
\begin{equation}
	ds_4^2=-\exp[-2\chi(r)]f(r)dt^2+\frac{dr^2}{f(r)}+r^2d\Omega^2_2,\quad \psi=\psi(r).\label{anstaz1}
\end{equation}

On substituting the ansatz (\ref{anstaz1}) to action $S_4$ in (\ref{action2}), we obtain the effective Lagrangian  
\begin{align}
	L_{\rm eff}=&e^{-\chi}\Big[2(1-\Lambda r^2 - f- r f') +\frac{2}{3}\Big(
	3 r^2 f^2 \psi '^3+2 r  \left(-r f'+2 r f \chi '-4 f\right)f \psi '^2-6  \Big(-r f'+2 r f \chi '\nonumber\\
	&-f+1\Big)f \psi '-6 (f-1) \left(f'-2 f \chi '\right)\Big)\alpha\psi'+ 4\alpha \lambda e^{-2\psi} \Big(r^2 f' \psi '-2 r^2 f \chi ' \psi '-3 r^2 f \psi '^2+r f'+f-1\Big)\nonumber\\
	&-6\alpha\lambda^2 r^2 e^{-4\psi} -2(1+2\gamma) q r^{2 \gamma} \Big]\,.
\end{align}
The dynamical equations for $f(r)$, $\chi(r)$, and field $\psi(r)$ are obtained from the Euler-Lagrange equations. We consider a special case of scalar field $\chi(r)=0$ \cite{Lu:2020iav}, these equations for the internally flat spacetime ($\lambda=0$), respectively, read as
\begin{eqnarray}
	&&\exp[\psi] \alpha \Big(1 -(1 - r \psi')^2f\Big) (\psi'^2+\psi'')=0,\label{A1}\\
	&&\exp[3\psi]\alpha\Bigg[ \Big(2\psi' + (1-r\psi')^2f'\Big)f' -f''- 2(1-r\psi') \left(-2\psi'^2+\psi''-3r\psi'\psi'' \right)f^2  + \Big((1-r\psi')^2f'' + 2\psi''\nonumber\\
	&&\qquad\;\;\;\;\;\;\; -2 (-1+r\psi' )f'\left(-3\psi' +2r\psi'^2-r \psi'' 
	\right)  \Big)f
	\Bigg]=0,\label{A2}\\
	&&\exp[3\psi]\Bigg[ 1 -(1+2\gamma) q r^{2 \gamma} -\Lambda r^2  -(r+2\alpha\psi')f' + \Bigg(-1+\alpha\psi' \Big(-2(1+f)\psi' +r^2f\psi'^3+2\Big(3+r\psi'(-3+r\psi' ) \Big)f'\Big)\nonumber\\
	&&\quad + 4\alpha\Big(-1+(-1+r\psi' )^2f\Big)\psi''\Bigg)f 
	\Bigg]=0.\label{A3}
\end{eqnarray}
Solving Eq.~(\ref{A1}), leads to the solution for the scalar field as follow:
\begin{equation}
	\psi(r)=\log\Big[\frac{r}{L}\Big] + \log[\cosh(\xi)-\sinh(\xi)], \quad \xi(r)=\int_{1}^{r}\frac{du}{u\sqrt{f(u)}},\label{A4}
\end{equation}
where $L$ is an integration constant. For the scalar field given in Eq.~(\ref{A4}), the dynamical equation for $\psi(r)$ in (\ref{A2}) is automatically satisfied, whereas Eq.~(\ref{A3}) yields the solution for metric function $f(r)$ as
\begin{equation}
	f_{\pm}(r)=1+\frac{r^2}{2\alpha}\left(1\pm\sqrt{1+\frac{8M\alpha}{r^3}+\frac{4\Lambda\alpha}{3} +\frac{4\alpha q}{r^{2(1-\gamma)}}}\right).\label{A9}
\end{equation}

Although this alternative approach of $4D$ regularization of EGB gravity is noteworthy different in spirit from the Glavan and Lin \cite{Glavan:2019inb}. Interestingly, these alternate regularization techniques \cite{Hennigar:2020lsl,Lu:2020iav} and Glavan and Lin's procedure \cite{Glavan:2019inb} lead to exactly same static spherically symmetric black hole solutions. Interestingly,  the solution (\ref{A9}) is exactly same as generated from our  theorems (see Table \ref{tab:table3}). However, a larger class of black hole solutions may exist in the $4D$ effective scalar-tensor gravity theory for the generic choices of scalar field ansatz \cite{Hennigar:2020lsl,Lu:2020iav,Ma:2020ufk}.

\begin{acknowledgments}
S.G.G would like to thanks DST INDO-SA bilateral project DST/INT/South Africa/P-06/2016 and SERB-DST for the ASEAN project IMRC/AISTDF/CRD/2018/000042. R.K. would like to thank UGC for providing SRF.
\end{acknowledgments}

\end{document}